\documentclass[showkeys,showpacs]{revtex4}

\usepackage{graphicx}
\usepackage{amsmath}
\usepackage{amssymb}

\begin{document}

\title{Spinor brane}

\author{
Vladimir Dzhunushaliev,$^{1,2,3}$
\footnote{
Email: vdzhunus@krsu.edu.kg}
Vladimir Folomeev~$^{2,3}$
\footnote{Email: vfolomeev@mail.ru}
}
\affiliation{$^1$Department of Physics and Microelectronic
Engineering, Kyrgyz-Russian Slavic University, Bishkek, Kievskaya Str.
44, 720021, Kyrgyz Republic \\ 
$^2$Institute of Physicotechnical Problems and Material Science of the NAS
of the
Kyrgyz Republic, 265 a, Chui Street, Bishkek, 720071,  Kyrgyz Republic \\
$^3$Institut f\"ur Physik, Universit\"at Oldenburg, Postfach 2503
D-26111 Oldenburg, Germany
}

\begin{abstract}
The thick brane model supported by a nonlinear spinor field is constructed. The different cases with the various values of the
cosmological constant
$\Lambda \left( \begin{smallmatrix}
< \\
= \\
>
\end{smallmatrix} \right)
0$
are investigated. It is shown that regular analytical spinor thick brane solutions with asymptotically Minkowski (at $\Lambda=0$)
or anti-de Sitter spacetimes (at $\Lambda<0$)  do exist.
\end{abstract}


\maketitle

\section{Introduction}
In the last years, the brane world models gained  widespread acceptance for a description of the Universe (for a review, see
\cite{Rubakov:2001kp,Barvinsky:2005ak}). In such models it is
assumed that our Universe is a four-dimensional hyperspace embedded in a higher-dimensional space.
Such an approach allows to solve some important problems of a theory of elementary particles. There are two types of
brane models: thin and thick ones. In the first case, it is assumed that a thickness of the brane is infinitely small, and matter
has delta-like distribution on the brane. However, from the physical point of view, it is obvious that  this is just some kind of idealization,
and more realistic models of the Universe demand introduction of the brane thickness. For this reason, a consideration of the
thick brane models  assumes ever greater importance (see e.g. the recent review \cite{Dzhunushaliev:2009va}).

Creation of  models of the thick branes is quite interesting  process by itself. 
The main difficulty consists in the necessity of obtaining a regular solution.
Here we  call the solution regular if the energy density per unit volume of the brane is finite. Usually, the thick brane solutions
 are being obtained as a result of the interaction of scalar fields with gravitational one. In this paper we present
a new model of the thick brane
supported by a nonlinear spinor field.

It is well-known \cite{finkelstein} that without gravitational field there exists a regular spherically symmetric solution for the spinor field. 
It is reasonable to suppose that the inclusion of the gravitational field does not destroy these solutions. However, self-consistent asymptotically flat solutions for a gravitating spinor field are not known until now. Note that the solutions for a spinor field propagating on a curved background are known (for details, see Ref. \cite{chandrasekhar}). The cosmological solutions with a spinor field are known as well, see Refs. \cite{Saha:2006iu}-\cite{Balantekin:2007km}. In Refs. \cite{Saha:2006iu}-\cite{Saha:2005ya}  the cosmological solutions with the gravitating spinor fields are considered. In Ref. \cite{Vakili:2005nb} the classical and quantum evolution of a universe in which the matter source is a massive Dirac spinor field and the universe is described by a Bianchi type I metric is investigated. It is shown that there exists the transition from an Euclidean to a Lorentzian regimes. In Ref. \cite{Balantekin:2007km}  a class of exact cosmological solutions with a neutral scalar field and  Majorana fermion field was found.

In this paper we consider a thick brane model supported by the nonlinear spinor field. The main purpose is to show that 
in the model with the nonlinear gravitating spinor field the existence of localized regular multidimensional solutions
trapping test matter fields is possible.

\section{5D brane from nonlinear spinor field}
\label{5Dbranespinor}

We consider a five-dimensional gravitation with the nonlinear spinor field as a source of matter.
Usually, in four-dimensional problems the following types of nonlinear terms are used:
$$
\left(\bar \psi \psi\right)^2, \quad \left(\bar \psi \gamma^5 \psi\right)^2, \quad
\left(\bar \psi \gamma^{\mu} \psi\right)\left(\bar \psi \gamma_{\mu} \psi\right), \quad
\left(\bar \psi \gamma^5 \gamma^{\mu} \psi\right)\left(\bar \psi \gamma^5 \gamma_{\mu} \psi\right).
$$
Here we choose the simplest variant when  the nonlinear term is taken in the form $\left(\bar \psi \psi\right)^2$.
Then the Lagrangian of the spinor field will be
\begin{equation}
 \mathcal L_m = \frac{i}{2} \left(
		\bar \psi \not \! \nabla \psi - \bar \psi \overleftarrow{\not \! \nabla} \psi
	\right) - m \bar \psi \psi  + \frac{\lambda}{2} \left(
		\bar \psi \psi
	\right)^2.
\label{2-40}
\end{equation}
The corresponding five-dimensional Einstein and Dirac equations are
\begin{eqnarray}
  R_a^{\phantom{a} A} - \frac{1}{2} e_a^{\phantom{a} A} R &=& \varkappa T_a^{\;A}
		+ e_a^{\phantom{a} A} \Lambda,
\label{2-10} \\
	\left[
	i \Gamma^a e_a^{\phantom{a} A}  D_A - m +
	\lambda \left( \bar \psi \psi \right)
	\right] \psi &=& 0,
\label{2-20}
\end{eqnarray}
where $a=\bar 0, \bar 1, \bar 2, \bar 3, \bar 5$ is the Lorentz index;
$A=0,1,2,3,5$ is the world index; $e_a^{\phantom{a}A}$ is the 5-bein; $\Gamma^a$
are the 5D Dirac matrices in a flat Minkowski space;
$D_A \psi = \left( \partial_A - \frac{1}{4} \omega_A^{\phantom{A} ab}
\Gamma_{ab} \right) \psi$ is the covariant derivative of the spinor $ \psi$;
$\Gamma_{ab} = \frac{1}{2}\left(\Gamma_a\Gamma_b-\Gamma_b\Gamma_a\right)$;
$ \not \! \nabla \psi=e^A_a \gamma^a D_A \psi$;
$m, \lambda$ are some parameters; $\Lambda$ is the cosmological constant. All definitions for the spinor differential geometry are taken from \cite{ortin}. The energy-momentum tensor for the spinor field is
\begin{equation}
\begin{split}
 T_a^{\;A } = &- \frac{i}{2} \bar \psi \left(
		\Gamma^A e_a^{\phantom{a} B} + \Gamma_a g^{AB}
	\right) D_B \psi + \frac{i}{2} D_B \bar \psi \left(
		\Gamma^A e_a^{\phantom{a} B} + \Gamma_a g^{AB}
	\right) \psi + e_a^{\phantom{a} A} \mathcal L_m -
\\
	&
	\frac{i}{2} \bar \psi \Gamma^A_{\phantom{A} a} \not{\! \nabla} \psi +
	\frac{i}{2} \bar \psi \overleftarrow{\not{\!\nabla}} \Gamma_a^{\phantom{a} A}
\psi,
\label{2-30}
\end{split}
\end{equation}
where the last two terms vanish on-shell; $\Gamma^A = e_a^{\,\, A} \Gamma^a$ are the
5D Dirac matrices in a curved spacetime;
$g^{AB} = e_a^{\phantom{a} A} e_b^{\,\, B} \eta^{ab}$ is the 5D contravariant metric tensor; $\eta^{ab} = \left\lbrace +1,-1,-1,-1,-1 \right\rbrace$ is the contravariant
metric tensor in the 5D Minkowski spacetime;
$\bar \psi = \bar \psi^\dagger \Gamma^{\bar 0}$ is the Dirac conjugated spinor;
$D_A \bar \psi = \bar \psi\left ( \overleftarrow \partial_A + \frac{1}{4}
\omega_A^{\phantom{A} ab} \Gamma_{ab} \right)$, where
$\bar \psi \overleftarrow \partial_A = \partial_A \bar \psi$.

The 5D Dirac matrices in a flat Minkowski space are
\begin{eqnarray}
  \Gamma^{\bar 0} &=& \begin{pmatrix}
		0												&		\mathbb I_{2 \times 2} \\
		\mathbb I_{2 \times 2} 	& 0
	\end{pmatrix},
\label{2-50}\\
  \Gamma^{\bar i} &=& \begin{pmatrix}
		0					&		-\sigma_{\bar i} \\
		\sigma_{\bar i}  & 0
	\end{pmatrix}, \bar i = 1,2,3,
\label{2-60} \\
  \Gamma^{\bar 5} &=& \begin{pmatrix}
		-i \mathbb I_{2 \times 2}	&		0 \\
		0   											& i \mathbb I_{2 \times 2}
	\end{pmatrix},
\label{2-70}
\end{eqnarray}
where $\mathbb I_{2 \times 2}$ is $2 \times 2$ unity matrix, and $\sigma_{\bar i}$
are Pauli matrixes
$$
  \sigma_{\bar 1} = \begin{pmatrix}
		0	 &		1 \\
		1  & 0
	\end{pmatrix}, \quad
\sigma_{\bar 2} = \begin{pmatrix}
		0	 & -i \\
		i  & 0
	\end{pmatrix}, \quad
\sigma_{\bar 3} = \begin{pmatrix}
		1	 &	0 \\
		0  & -1
	\end{pmatrix}.
$$

We seek a wall-like solution for the system \eqref{2-10}-\eqref{2-20}. To do this let us choose the 5D bulk metric in the form
\begin{equation}
 ds^2 = \phi^2 (r) ds^2_M - dr^2,
\label{2-110}
\end{equation}
where $ ds^2_M$ is the 4D Minkowski metric. For the spinor field, we use the following ansatz
\begin{equation}
 \psi = \begin{pmatrix}
		a(r) 	\\
		0 		\\
		b(r)	\\
		0
	\end{pmatrix}.
\label{2-115}
\end{equation}
Then, using Eqs. \eqref{2-10} and \eqref{2-20}, one can obtain
\begin{eqnarray}
  \frac{\phi ''}{\phi} + \frac{{\phi'}^2}{\phi^2} &=&
	\frac{2}{3} \varkappa \lambda\, a^2 b^2 - \frac{\Lambda}{3} ,
\label{2-120} \\
	\frac{{\phi'}^2}{\phi^2} &=& \frac{\varkappa}{3} \left[
		\frac{1}{2}\left(a' b - a b'\right) + \lambda a^2 b^2
	\right] - \frac{\Lambda}{6},
\label{2-130} \\
  a' + 2\frac{\phi'}{\phi} a - m a +
	2 \lambda a^2 b &=&	0 ,
\label{2-140} \\
	b' +2\frac{\phi'}{\phi} b  + m b -
	2 \lambda a b^2 &=&	0 .
\label{2-150}
\end{eqnarray}
Introducing new dimensionless variables
$\tilde{r} = m r$, $\tilde{a}=a \sqrt{2 \lambda/m} $,
$\tilde{b}=b \sqrt{2 \lambda/m} $,
 $\tilde{\varkappa}=\varkappa/(4\lambda)$, $\tilde{\Lambda}=\Lambda/m^2$, and using Eqs. \eqref{2-140} and \eqref{2-150}  for eliminating derivatives on the right hand side of  Eq. \eqref{2-130}, we have
\begin{eqnarray}
  \frac{\phi ''}{\phi} + \frac{{\phi'}^2}{\phi^2} &=&
\frac{2}{3}\tilde{\varkappa} \tilde{a}^2 \tilde{b}^2 - \frac{\tilde{\Lambda}}{3} ,
\label{2-160} \\
	\frac{{\phi'}^2}{\phi^2} &=& \frac{\tilde{\varkappa}}{3} \left(2 \tilde{a} \tilde{b}-\tilde{a}^2 \tilde{b}^2\right) - \frac{\tilde{\Lambda}}{6},
\label{2-170} \\
  \tilde{a}' + 2\frac{\phi'}{\phi} \tilde{a} -  \tilde{a} +
	 \tilde{a}^2 \tilde{b} &=&	0 ,
\label{2-180} \\
\tilde{b}' +2\frac{\phi'}{\phi} \tilde{b}  +  \tilde{b} -
	 \tilde{a} \tilde{b}^2 &=&	0 .
\label{2-190}
\end{eqnarray}
This system  can be solved analytically. To do this let us find from Eqs.
 \eqref{2-180} and \eqref{2-190} the following first integral
\begin{equation}
\tilde{a}\tilde{b} \phi^4=C,
\label{relation}
\end{equation}
where $C$ is an integration constant. Then Eq. \eqref{2-170} can be easily integrated giving three types of solutions:

1) At $\tilde{\Lambda}=0$. In this case
\begin{eqnarray}
\label{phi_sol_0}
		\phi &=& \left[
		\frac{1}{2}C\left(1+\frac{16}{3}\tilde{\varkappa} \tilde{r}^2\right)
		\right]^{1/4},
\\
\label{ab_sol_0}
	a &=& a_0\exp\left\{
	\tilde{r}-\frac{1}{2}\left[
		\sqrt{\frac{3}{\tilde{\varkappa}}}
		\arctan\left(4\sqrt{\frac{\tilde{\varkappa}}{3}}\;\;\tilde{r}\right)
		+\ln\left(1+\frac{16}{3}\tilde{\varkappa}\;\tilde{r}^2\right)
		\right]
	\right\},
\\
\label{ab_sol_1}
	b &=& b_0\exp\left\{
	-\tilde{r}+\frac{1}{2}\left[
		\sqrt{\frac{3}{\tilde{\varkappa}}}
		\arctan\left(4\sqrt{\frac{\tilde{\varkappa}}{3}}\;\;\tilde{r}\right)
		-\ln\left(1+\frac{16}{3}\tilde{\varkappa}\;\tilde{r}^2\right)
		\right]
	\right\},
\end{eqnarray}
where $a_0, b_0$ are integration constants. In the case under consideration $a_0=b_0=\sqrt{C}$.

2) At $\tilde{\Lambda}>0$. Here Eq. \eqref{2-170} gives
\begin{equation}
	\phi_{1,2}=\left[\frac{1}{2}\left(p \pm \sqrt{p^2-
	\frac{4 q A^2+p^2}{A^2+1}}\right)\right]^{1/4}
\label{phi_sol_plus}
\end{equation}
with
$$
	A=\tan{\left(
	\mp 4\sqrt{\frac{\tilde{\Lambda}}{6}}(\tilde{r}+\tilde{r}_0)
	\right)}, \quad
	p=\frac{4\tilde{\varkappa} C}{\tilde{\Lambda}},
	\quad q=\frac{2\tilde{\varkappa} C^2}{\tilde{\Lambda}},
$$
where $\tilde{r}_0$ is an integration constant. In this case, it is also possible to find analytical solutions for $a$ and $b$ but they are quite tedious, and that is why we do not show them here.

3) At $\tilde{\Lambda}<0$. In this case we have
\begin{eqnarray}
\label{phi_sol_minus}
	\phi &=& \left\{\frac{1}{4B}\left[(B-p)^2+4q\right]\right\}^{1/4},
\\
\label{ab_sol_minus}
	a &=& a_0 \exp\left\{	\tilde{r}-
	\frac{1}{2}C\sqrt{\frac{6}{q|\tilde{\Lambda}|}}\arctan\left[
		\frac{1}{2\sqrt{q}}(B-p)
	\right]-
	\frac{1}{2}\ln\left[\frac{(B-p)^2+4q}{4B}\right]
	\right\},
\\
	b &=& b_0 \exp\left\{
	-\tilde{r}+\frac{1}{2}C\sqrt{\frac{6}{q|\tilde{\Lambda}|}}\arctan\left[
		\frac{1}{2\sqrt{q}}(B-p)
	\right]-
	\frac{1}{2}\ln\left[\frac{(B-p)^2+4q}{4B}\right]
	\right\}
\end{eqnarray}
with
$$
	B=\gamma \exp{\left(4\sqrt{\frac{|\tilde{\Lambda}|}{6}}
	\,\,\,|\tilde{r}|\right)}, \quad
	p=\frac{4\tilde{\varkappa} C}{|\tilde{\Lambda}|},
	\quad q=\frac{2\tilde{\varkappa} C^2}{|\tilde{\Lambda}|},
$$
where $a_0, b_0, \gamma$ are integration constants.

The corresponding dimensionless energy density is
$$
	\tilde{T}_0^0=\frac{\varkappa}{m^2} T^0_0=
	-2\tilde{\varkappa}  \tilde{a}^2\tilde{b}^2.
$$
One can see from Eqs. \eqref{relation}-\eqref{ab_sol_minus} that for
 $\Lambda=0$   and  $\Lambda<0$ the energy integrals on the coordinate $\tilde{r}$ are finite
\begin{equation}
\label{energy}
	\int \limits_{-\infty}^{\infty} \tilde{T}^0_0 \sqrt{-^5g} \,d\tilde{r}=
	-2\tilde{\varkappa} C^2
	\int \limits_{-\infty}^{\infty}\frac{d\tilde{r}}{\phi^4}<\infty,
\end{equation}
i.e. the solutions are regular  and one can expect that such solutions can be used for modeling the branes. It is just necessary to check up whether test fields will be trapped  by these solutions (see a consideration of this question in the next section). On the other hand, in the case when $\Lambda>0$ this integral diverges, and that is why using such solution for modeling a brane is not straightforward (see discussion of this question in the Conclusion).

\begin{figure}
  \includegraphics[width=10cm]{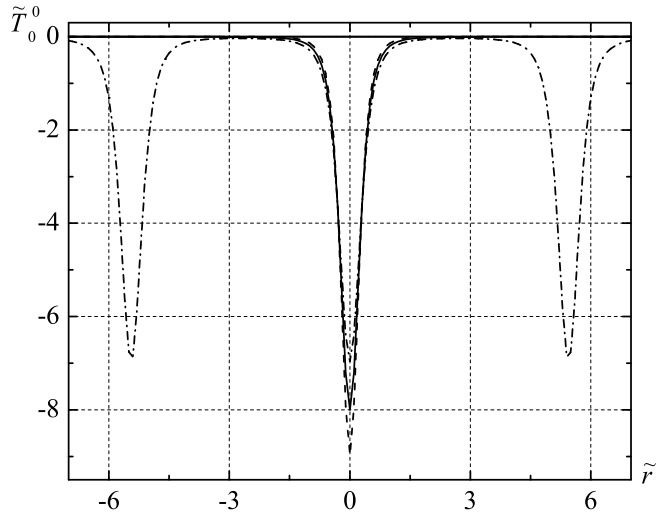}
	\vspace{-1cm}
	  \caption{The profiles of the energy density $\tilde{T}^0_0$
	  for $\tilde{\Lambda}=0$ (the solid line),
	  $\tilde{\Lambda}=-0.5$ (the dashed line),
	  $\tilde{\Lambda}=0.5$ (the dash-dotted line),
	  $\tilde{\varkappa}=1$. The profiles for the curves $\tilde{\Lambda}=0$ and
	  $\tilde{\Lambda}=0.5$ practically coincide.}
    \label{energy_f}
\end{figure}

To demonstrate the behavior of the solutions, let us solve the system \eqref{2-160}-\eqref{2-190} with the following boundary conditions
\begin{equation}
  \phi(0) = 1, \quad
	\phi'(0) = 0, \quad
	\tilde{a}(0) = \tilde{b}(0)=\left(
	1+\sqrt{1-\frac{\tilde{\Lambda}}{2\tilde{\varkappa}}}\,\,
	\right)^{1/2}.
\label{2-200}
\end{equation}
The last condition follows from the constraint equation \eqref{2-170}. We set $\phi(0) = 1$ because we can redefine the Minkowski coordinates $x^\mu$ arbitrarily, $\phi'(0) = 0$ because the bulk space should be symmetrical relative to the origin of coordinates $\tilde{r} = 0$. The spinor \eqref{2-115} had been chosen in the spinor representation, and in the standard representation the spinor components have to be either odd or even functions. That is why we choose
$\tilde{a}(0) = \tilde{b}(0)$. The corresponding results for the energy density \eqref{energy} for three different values of $\Lambda$ are presented in Fig.~\ref{energy_f}. The behavior of the metric function $\phi$ and the spinor functions $a,b$ can be estimated from the analytical expressions
obtained above.

\section{Trapping of matter}

The five-dimensional localized wall-like solutions obtained above can be used for a description of a brane only if it will be possible to show that various test matter  fields can be confined on such wall. As an example of such field, let us consider here a test complex scalar filed $\chi$ with the Lagrangian
$$
	L_{\chi}=\frac{1}{2}\partial_A \chi^{*}\partial^A\chi-
	\frac{1}{2}m_0^2 \chi^*\chi,
$$
where $m_0$ is the mass of the test field. Using this Lagrangian, we find the equation for the scalar field
\begin{equation}
\label{tf_eq}
	\frac{1}{\sqrt{-^5 g}}\frac{\partial}{\partial x^A}\left(
		\sqrt{-^5 g} g^{AB}\frac{\partial \chi}{\partial x^B}
	\right)=-m_0^2 \chi.
\end{equation}
Here $\sqrt{-^5 g}$ is the determinant of the 5D metric $g_{AB}$ and $\chi$ is a function of all coordinates $\chi=\chi(x^A)$. Taking into account that the  canonically conjugate momenta $p_\mu=(E, \overrightarrow{p})$ are integrals of motion, we seek a solution in the form
$$
  \chi (x^{A}) = X(\tilde{r}) \exp (-ip_{\mu }x^{\mu }).
$$
Substituting this ansatz in  \eqref{tf_eq}, one can find the following equation for $X(\tilde{r})$
$$
    X^{\prime\prime} + 4\frac{\phi^{\prime}}{\phi}X^{\prime}+ (p^{\mu }p_{\mu } -m_{0}^2 ) X = 0,
$$
or, taking into account that
$p^{\mu}p_{\mu}= \phi^{-2} \left(E^2-\overrightarrow{p}^2\right)$, we have
\begin{equation}
\label{test_eq}
	X^{\prime\prime} +4\frac{\phi^{\prime}}{\phi}X^{\prime}+ \left[
	\left(E^2-\overrightarrow{p}^2\right)\phi^{-2}-m_0^2
	\right] X = 0,
\end{equation}
where  the prime denotes differentiation with respect to $\tilde{r}$.
In the case when $\tilde{\Lambda}=0$, the asymptotic behavior of Eq.~\eqref{phi_sol_0} is $\phi\approx \beta \sqrt{|\tilde{r}|}$, where
$\beta=\left(\frac{8}{3}C\tilde{\varkappa}\right)^{1/4}>0$.  Then we have from \eqref{test_eq}
$$
	X^{\prime\prime}+\frac{2}{\tilde{r}}X^{\prime}-m_0^2  X = 0.
$$
This equation has an asymptotically decaying solution in the form
\begin{equation}
\label{asymp_sol_0}
	X_{\infty}^{(0)}\approx D \frac{e^{-m_0 |\tilde{r}|}}{|\tilde{r}|},
\end{equation}
where $D$ is an integration constant and the index $(0)$ refers to the solutions with $\tilde{\Lambda}=0$.

In the case when $\tilde{\Lambda}<0$, the asymptotic follows from Eq. \eqref{phi_sol_minus}:  $\phi\approx\left(\frac{1}{4} B\right)^{1/4}$.
Then Eq.~\eqref{test_eq} gives
$$
		X^{\prime\prime}+4\sqrt{\frac{|\tilde{\Lambda}|}{6}}X^{\prime}-m_0^2  X = 0,
$$
with an asymptotically decaying solution
\begin{equation}
\label{asymp_sol_minus}
	X_{\infty}^{(-)}\approx D \exp{\left[-2\left(\sqrt{\frac{|\tilde{\Lambda}|}{6}}+
\sqrt{\frac{|\tilde{\Lambda}|}{6}+\frac{m_0^2}{4}}\,\,\right)|\tilde{r}|\right]}
\end{equation}
where $D$ is an integration constant
and the index $(-)$ refers to the solutions with $\tilde{\Lambda}<0$.

As a necessary condition of trapping of the matter, one can require converging the field energy per unit 3-volume of the brane \cite{Abdyrakhmanov:2005fs}, i.e.
\begin{equation}
\label{E_tot}
     E_{\rm tot}[\chi] = \int \limits_{-\infty}^{\infty} T^0_0 \sqrt{-^5g} \,d\tilde{r}
          = \int\limits_{-\infty}^{\infty}
            \phi^4\left[ \frac{1}{\phi^2}(E^2 + \overrightarrow{p}^2)X^2
	    + m_0^2 X^2 + X^{\prime 2} \right] d\tilde{r} < \infty,
\end{equation}
and also the norm of the field $\chi$ should be finite
$$
	||\chi||^2 = \int\limits_{-\infty }^{\infty }\sqrt{-^5g}\,\chi^*\chi\,d\tilde{r}
    	       = \int_{-\infty }^{\infty } \phi^4\,X^2 \,d\tilde{r}.
$$
Taking into account the asymptotic solutions \eqref{asymp_sol_0} and \eqref{asymp_sol_minus} for $\tilde{\Lambda}=0$ and $\tilde{\Lambda}<0$, respectively, one can see that $E_{\rm tot}$ and $||\chi||$ converge asymptotically for both cases.
Thus, it is obvious from the above analysis that the localized solutions which we found in section \ref{5Dbranespinor} trap the test scalar field
 that indicates that such solutions may be interpreted as brane solutions.

\section{Conclusion}
In this paper, we have considered  the $Z_2$-symmetric domain wall (thick brane)
solutions supported by the
nonlinear spinor field both with and without account of the five-dimensional
cosmological $\Lambda$-term. It was shown that regular solutions, i.e. solutions with the finite
energy density per unit 3-volume of the brane, do exist. In the case of
positive  $\Lambda$ one can see that the total energy density over all space diverges.
In the cases  $\Lambda=0$ and  $\Lambda<0$ the total energy density remains finite. It allows
 using such solutions for a description of the thick branes. To show this, we have considered  trapping
 of the test scalar field on the 	wall. It was shown that asymptotically converging solutions exist
for the test scalar field both for  $\Lambda=0$ and  $\Lambda<0$ cases.
The asymptotic behavior of the brane solutions corresponds to  the Minkowski (at $\Lambda=0$)
or anti-de Sitter spacetimes (at $\Lambda<0$).

As regards the case $\Lambda>0$, the obtained periodic solutions for the energy density (see Fig.~\ref{energy_f}) can be used for a description of the five-dimensional spacetime with the compactified fifth coordinate. In this case the size of the fifth dimension is defined by the period of the obtained solution for the spinor field.


\section*{Acknowledgements}

V.D. is grateful to the Research Group Linkage Programme of the Alexander von Humboldt Foundation for the support of this research. V.F. would like to thank the German Academic Exchange Service  (DAAD) for financial support. We  would like to express our gratitude to the Department of Physics of the Carl von Ossietzky University of Oldenburg  and, specially, to Prof. J. Kunz and Prof. B. Kleihaus for fruitful discussions.


\begin{thebibliography}{99}
\bibitem{Rubakov:2001kp}
  V.~A.~Rubakov,
  Phys.\ Usp.\  {\bf 44}, 871 (2001)
  [Usp.\ Fiz.\ Nauk {\bf 171}, 913 (2001)]
  [arXiv:hep-ph/0104152].
\bibitem{Barvinsky:2005ak}
  A.~O.~Barvinsky,
  Phys.\ Usp.\  {\bf 48}, 545 (2005)
  [Usp.\ Fiz.\ Nauk {\bf 175}, 569 (2005)].

\bibitem{Dzhunushaliev:2009va}
  V.~Dzhunushaliev, V.~Folomeev and M.~Minamitsuji,
  arXiv:0904.1775 [gr-qc].
\bibitem{finkelstein}
R. Finkelstein, R.Lelevier and M. Ruderman,
Phys. Rev., {\bf 83}, 326(1951); \\
R. Finkelstein, C. Fronsdal and P.Kaus,
Phys. Rev., {\bf 103}, 1571(1956).

\bibitem{chandrasekhar}
S. Chandrasekhar,
``The mathematical theory of black holes'',
Clarendon Press Oxford, Oxford University Press New York, 1983.

\bibitem{Saha:2006iu}
  B.~Saha,
  Phys.\ Rev.\  D {\bf 74}, 124030 (2006).

\bibitem{Saha:2003xv}
  B.~Saha and T.~Boyadjiev,
  Phys.\ Rev.\  D {\bf 69}, 124010 (2004)
  [arXiv:gr-qc/0311045].

\bibitem{Saha:2005ya}
  B.~Saha,
  Grav.\ Cosmol.\  {\bf 12}, 215 (2006)
  [arXiv:gr-qc/0512050].

\bibitem{Vakili:2005nb}
  B.~Vakili and H.~R.~Sepangi,
  JCAP {\bf 0509}, 008 (2005)
  [arXiv:gr-qc/0508090].

\bibitem{Balantekin:2007km}
  A.~B.~Balantekin and T.~Dereli,
  Phys.\ Rev.\  D {\bf 75}, 024039 (2007)
  [arXiv:gr-qc/0701025].

\bibitem{ortin}
T. Ortin,
``Gravity and Strings'',
Cambridge, UK ; New York : Cambridge University Press, 2004.

\bibitem{Abdyrakhmanov:2005fs}
  S.~T.~Abdyrakhmanov, K.~A.~Bronnikov and B.~E.~Meierovich,
  Grav.\ Cosmol.\  {\bf 11}, 82 (2005)
  [arXiv:gr-qc/0503055].



\end{thebibliography}
\end{document}